\documentclass[]{article}
\usepackage[]{graphicx}
\usepackage{amsmath}
\usepackage{amsfonts}
\usepackage{amsthm}
\usepackage{amstext}
\usepackage{amssymb}
\usepackage{amscd}

\pdfoutput=1

\def \khat{\boldsymbol{\hat{k}}}%
\def \nhat{\boldsymbol{\hat{n}}}%
\def \what{\boldsymbol{\hat{w}}}%
\def \r{\boldsymbol{r}}%
\def \v{\boldsymbol{v}}%
\def \w{\boldsymbol{w}}%

\def \A{\boldsymbol{A}}%
\def \J{\boldsymbol{J}}%

\begin{document}
\title{Solutions of New Potential Integral Equations Using MLFMA Based on the Approximate Stable Diagonalization}

\author{{
U\u{g}ur Meri\c{c} G\"{u}r and \"{O}zg\"{u}r Erg\"{u}l}\\
Department of Electrical and Electronics Engineering\\ 
Middle East Technical University, Ankara, Turkey\\
ozergul@metu.edu.tr\vspace{0.3cm}\\
\small{Original Version: August, 2017}
}

\maketitle

\begin{abstract}
We present efficient solutions of recently developed potential integral equations~(PIEs) using a low-frequency implementation of the multilevel fast multipole algorithm (MLFMA). PIEs enable accurate solutions of low-frequency problems involving small objects and/or small discretization elements with respect to wavelength. As the number of unknowns grows, however, PIEs need to be solved via fast algorithms, which are also tolerant to low-frequency breakdowns. Using an approximate diagonalization in MLFMA, we present a new implementation that can provide accurate, stable, and efficient solutions of low-frequency problems involving large numbers of unknowns. The effectiveness of the implementation is demonstrated on canonical problems.                            
\end{abstract}

\section{Introduction}
In computational electromagnetics, low-frequency problems are known as those involving small objects and/or discretizations with respect to wavelength. In the literature of the last several decades, there is a collective effort to efficiently and accurately solve low-frequency problems~\cite{zhao2000_2}--\cite{ic120}, considering that conventional formulations and solution algorithms have numerical breakdowns that limit their applicability to various important problems in the areas of antennas, microwaves, and optics. It is remarkable that a low-frequency breakdown can be experienced as inaccuracy, inefficiency, and/or instability, not only for small-scale objects but also for dense discretizations that may be required for accurate analysis of many structures in diverse applications.\\
\indent Surface integral equations are widely used in numerical solutions of electromagnetic problems involving perfectly conducting objects. Among different kinds, the electric-field integral equation~(EFIE) is often preferred since it can be used for both open and closed surfaces. Unfortunately, EFIE suffers from a low-frequency breakdown~\cite{vipiana},\cite{andriulli2} due to its unbalanced terms (specifically, the dominating scalar potential part besides a vanishing vector potential part) when using the conventional discretizations. So far, this breakdown is mitigated by various methods~\cite{aefie},\cite{andriulli},\cite{yan},\cite{andriulli2}, all targeting separation/extraction of the electric charge density from the electric current density or rescaling the associated terms to capture the low-frequency physics. As a recent contribution, potential integral equations~(PIEs)~\cite{chew},\cite{greengard},\cite{jp063} are proposed to overcome the low-frequency breakdown of surface integral equations for both open and closed surfaces. As a major advantage, PIEs can be discretized with the conventional functions, such as the Rao-Wilton-Glisson~(RWG) functions on triangulated surfaces, without resorting to special discretization schemes.\\
\indent It has been shown that PIEs can provide accurate and stable solutions of electromagnetic problems involving very small objects and discretizations with respect to wavelength~\cite{chew},\cite{greengard},\cite{jp063}. On the other hand, these demonstrations have been limited to small numbers of unknowns that can easily be handled via the method of moments~(MOM). As the problem size grows, acceleration methods are needed, while they should also be resistant to low-frequency breakdowns themselves. In this contribution, we show the solution of PIEs using the multilevel fast multipole algorithm~(MLFMA)~\cite{zhao2000_2} based on the approximate diagonalization of the Green's function~\cite{jp046}. This type of diagonalization was used for the magnetic-field integral equation~(MFIE) for closed surfaces, while we now demonstrate its applicability to PIEs, leading to accurate, stable, and also efficient implementations to solve densely discretized objects at arbitrarily low frequencies. Initial results of this study involving small-scale objects was presented in~\cite{ic120}.             

\section{Potential Integral Equations}
We consider perfectly conducting objects located in free space with permittivity $\epsilon_{0}$ and permeability $\mu_{0}$. For a given object with surface $S$, recently developed potential formulations can be written as two coupled integral equations (in the frequency domain using $\exp(-i\omega t)$ time convention) as 
\begin{alignat}{1}
\nhat \times \mu_{0} \int_{S}{d\r' \J(\r')}g(\r,\r')&+\nhat \times \int_{S}{d\r' \nhat' \cdot \A(\r') \nabla'g(\r,\r')}\notag  \\ &=-\nhat \times\A^{\text{inc}}(\r) \quad (\r \in S) \label{ie2}\\
\int_{S}{d\r' \nabla'\cdot \J(\r')g(\r,\r')}&+\omega^{2}\epsilon_{0} \int_{S}{d\r' \nhat' \cdot \A(\r') g(\r,\r')}\notag \\ &=-i\omega\epsilon_{0}\Phi^{\text{inc}}(\r) \quad (\r \in S),\label{ie3}
\end{alignat} 
where $\A^{\text{inc}}$ and $\Phi^{\text{inc}}$ are the incident magnetic vector potential and the incident electric scalar potential, respectively. In the above, $\nhat$ is the unit normal to the surface, $\J$ is the induced electric current density, $\A$ is the total magnetic vector potential, and $g(\r,\r')=\exp(ik_{0}|\r-\r'|)/(4\pi|\r-\r'|)$ is the free-space Green's function with wavenumber $k_{0}=\omega\sqrt{\mu_{0}\epsilon_{0}}$. Equations (\ref{ie2}) and (\ref{ie3}) are solved to obtain the unknowns $\J$ and $\nhat' \cdot \A$. For far-zone scattering, the electric current density $\J$ is sufficient to compute the electric field intensity and the magnetic field intensity, while the near-zone electric field intensity can be calculated by finding the electric charge density as $\rho=i\omega\epsilon_{0}\nhat'\cdot \A-\epsilon_{0} \nhat'\cdot \nabla' \Phi$, where the gradient of the scalar potential is obtained by solving~\cite{jp063} 
\begin{alignat}{1}
\int_{S}{d\r' \nhat' \cdot \nabla'\Phi(\r') g(\r,\r')}=\Phi^{\text{inc}}(\r) \quad (\r \in S).
\end{alignat}
For numerical solutions, $\J$ and $\nhat'\cdot \A$ are expanded in terms of the RWG functions (defined on pairs of triangles) and pulse functions (defined on triangles), respectively. Then, (\ref{ie2}) and (\ref{ie3}) are tested by using the same sets of the RWG/pulse functions to construct $N\times N$ matrix equations with $N=T+E$, where $T$ is the number of the pulse functions (triangles) and $E$ is the number of the RWG functions (edges). PIEs are stable at arbitrarily low frequencies and for dense triangulations. As opposed to MFIE, which is also tolerant to low-frequency breakdowns, PIEs can be used for both open and closed geometries, making them useful for practical problems. 

\section{Approximate Diagonalization of the Green's Function}
Iterative solutions of PIEs directly via MOM need ${\cal O}(N^{2})$ memory and ${\cal O}(N^{2})$ time per matrix-vector multiplication~(MVM). Therefore, as the problem size grows in terms of the number of unknowns ($N$), fast solvers, such as MLFMA, are needed to perform efficient solutions with limited computational resources. However, most of the fast algorithms also have low-frequency breakdowns that inhibit their application to dense discretizations with respect to wavelength. In the context of MLFMA, the conventional plane-wave expansion fails to provide accurate computations of interactions for short distances~\cite{zhao2000_2},\cite{ergul2014_1}. Therefore, low-frequency and broadband implementations of MLFMA have been developed by using multipoles (directly without diagonalization) or by resorting to alternative integration schemes incorporating evanescent waves~\cite{jiang2004}--\cite{wallen2007},\cite{bogaert2008},\cite{vikram2009}. As an alternative to these elaborate schemes, an approximate diagonalization~\cite{jp045},\cite{jp046}, which does not need a complete re-implementation of MLFMA, is proposed recently for low-frequency and multi-scale applications. This technique has been used to solve challenging problems formulated with MFIE, while in this study, it is used for the first time to solve low-frequency problems formulated with PIEs.     

\begin{figure}[t!]
\begin{center}
\includegraphics[width=8.8cm]{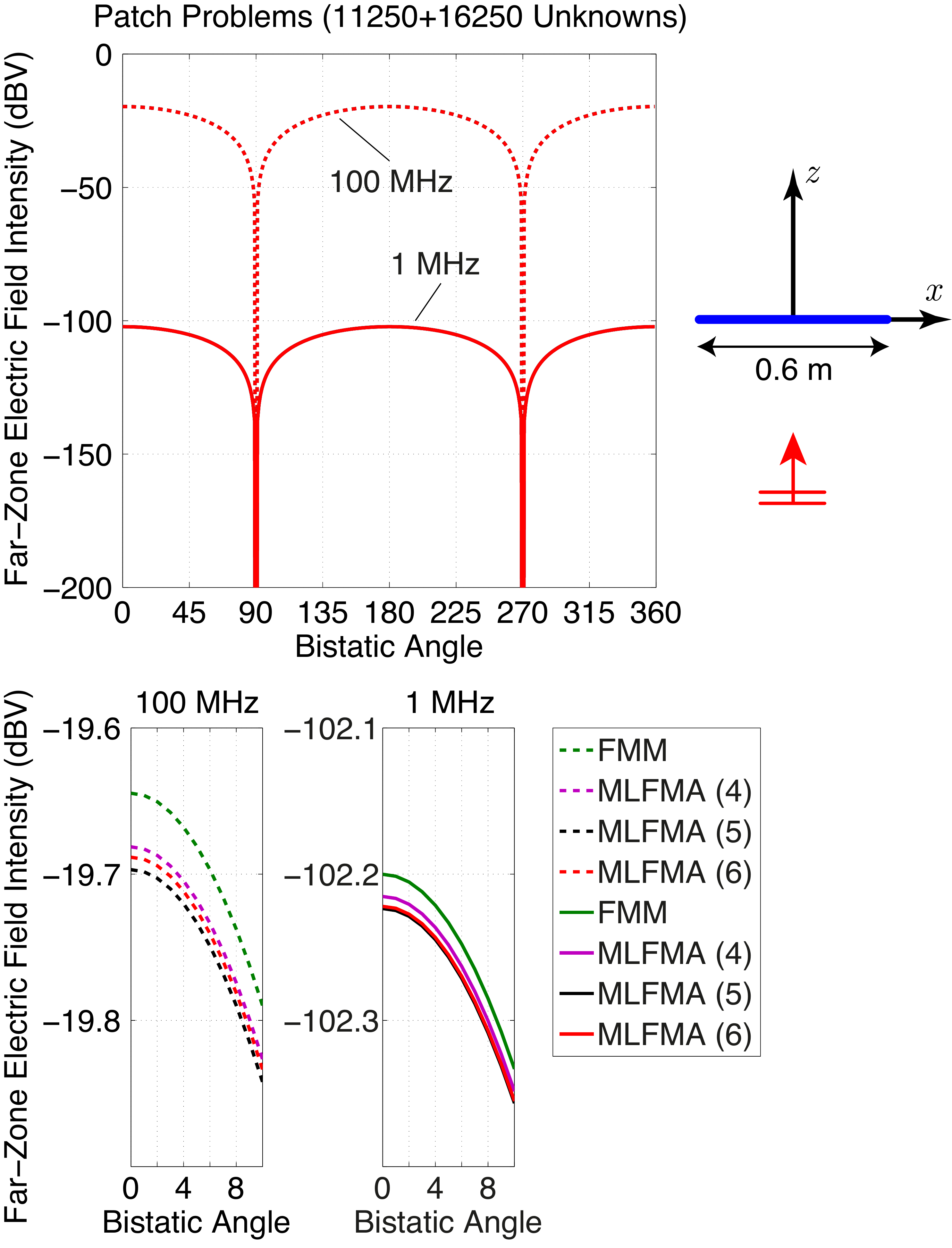}
\end{center}
\vspace{-0.3cm}
\caption{Solutions of scattering problems involving a metallic patch (edge length: $0.6$~m) at $1$~MHz and $100$~MHz. MLFMA  with different numbers of levels is used.}\label{fig001}\vspace{-0.2cm}
\end{figure}

According to the approximate diagonalization, the Green's function is written as an angular integration as  
\begin{alignat}{1}
g(\r,\r')&\approx \frac{ik_{0}}{4\pi}\int{d^{2}\khat \exp(ik_{0}\khat \cdot \v/s)\tilde{\alpha}(k_{0},\khat,\w)}
\end{alignat}
in terms of the scaled plane waves $\exp(ik_{0}\khat \cdot \v/s)$ and modified translation functions $\tilde{\alpha}(k_{0},\khat,\w)$, where $\r-\r'=\w+\v$ and $s$ is a scale factor. We further have  
\begin{alignat}{1}
\tilde{\alpha}(k,\khat,\w)&\approx\sum_{t=0}^{\tau}{i^{t}(2t+1)s^{t}h_{t}^{(1)}(kw)P_{t}(\khat \cdot \what)},
\end{alignat}
where $\tau$ is the truncation number, $P_{t}$ represents Legendre polynomials, and $h_{t}^{(1)}$ represents spherical Hankel functions. In the implementation, we use a standard clustering by enclosing the object in a cubic box and recursively dividing it into subboxes. This way, in each MVM, electromagnetic interactions are computed as a sequence of aggregation, translation, and disaggregation stages. Optimal values of the truncation numbers $\tau$ and the scaling factors $s$ are found and tabulated~\cite{jp045} for different box sizes with respect to wavelength while using the one-box-buffer scheme. The approximate diagonalization is directly applicable to PIEs since the differential operator on the Green's function in (\ref{ie2}) can be placed on the testing functions via a practice of integration by parts so that the Green's function remains as the only kernel of PIEs.       

\begin{figure}[t!]
\begin{center}
\includegraphics[width=8.8cm]{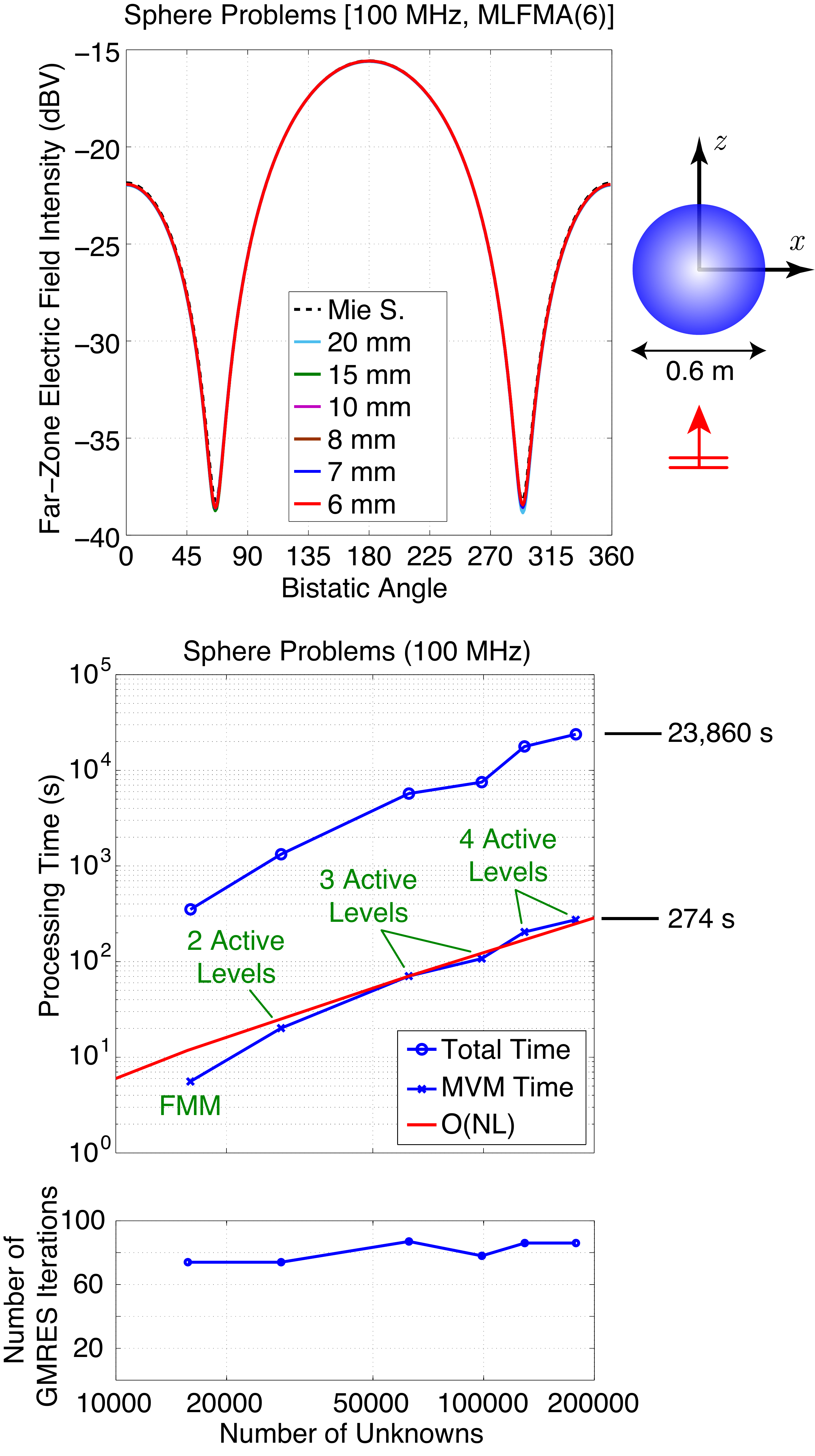}
\end{center}
\vspace{-0.3cm}
\caption{Solutions of scattering problems involving a metallic sphere (diameter: $0.6$~m) at $100$~MHz. MLFMA is used for different mesh sizes and number of unknowns.}\label{fig002}\vspace{-0.2cm}
\end{figure}

\section{Numerical Results and Conclusions}
As numerical examples, we consider scattering problems involving a perfectly conducting patch (with $0.6$~m edges) and a perfectly conducting sphere (with $0.6$~m diameter) located in free space. Both objects are at the origin and illuminated by plane waves propagating in the $z$ direction with $x$-polarized electric field. Fig.~\ref{fig001} presents the solutions of the patch problems when the frequency is $1$~MHz and $100$~MHz. We note that the edges of the patch correspond to $\lambda/500$ and $\lambda/5$, respectively, where $\lambda$ is the wavelength at these frequencies. For numerical solutions, the patch is discretized with $11250$ pulse and $16250$ RWG functions. Fig.~\ref{fig001} depicts the far-zone electric field intensity on the $z$-$x$ plane with respect to the bistatic angle $\theta$, where $0^{\circ}$ and  $180^{\circ}$ correspond to forwardscattering and backscattering, respectively. Solutions are obtained by using MLFMA with different number of levels, i.e., $3$, $4$, $5$, and $6$, where using three levels corresponds to a translation at a single level (FMM: fast multipole method). We observe that all numerical results are on the top of each other when using $200$~dB dynamic range. Since there is no direct reference (e.g., analytical) result, we show the consistency of the results in separated zoomed plots with $0.3$~dB dynamic ranges. The results demonstrate the excellent accuracy and stability of the implementation based on PIEs and MLFMA.

\indent Fig.~\ref{fig002} presents the results of the sphere problems at $100$~MHz (diameter:~$\lambda/5$) when the object is discretized with triangles of various sizes from $20$~mm~($\lambda/150$) to $6$~mm~($\lambda/500$). The far-zone electric field intensity, obtained by using six-level MLFMA for different mesh sizes, is plotted with respect to the bistatic angle on the $z$-$x$ plane. We observe that all numerical results are consistent with each other, and also with analytical results obtained via a Mie-series solution. As complementary results, Fig.~\ref{fig002} also presents the processing time measured on a single core of an E5-2680v3 processor in the MATLAB environment, as well as the number of generalized-minimal residual (GMRES) iterations with respect to the number of unknowns ($N$). For the mesh sizes from $20$~mm to $6$~mm, the number of unknowns is in the range from $15690$ to $178160$. To demonstrate the time complexity, MLFMA is used with different numbers of levels as the problem size grows. The processing time for an MVM changes from $34.8$~s ($15690$ unknowns) to $274$~s ($178160$ unknowns). For the total solution time, these values are $2611$~s ($15690$ unknowns) and $23840$~s ($178160$ unknowns), respectively. An ${\cal O}(NL)$ curve, where $N$ is the number of unknowns and $L$ is the number of active MLFMA levels (levels at which translations occur), is also shown in Fig.~\ref{fig002} to demonstrate the linearithmic complexity of MVMs. We note that the number of iterations increases slightly from $74$ to $86$, while the number of unknowns increases more than $11$ times. As demonstrated in these numerical results, we conclude that an implementation based on PIEs and MLFMA using the approximate diagonalization provides accurate, stable, and efficient solutions of low-frequency problems involving small elements and large numbers of unknowns.

\section*{Acknowledgment}This work was supported by the Scientific and Technical Research Council of Turkey (TUBITAK) under the Research Grant 114E498 and by the Turkish Academy of Sciences (TUBA).

\end{document}